\documentclass[sn-mathphys-num]{sn-jnl}

\usepackage{graphicx}%
\usepackage{multirow}%
\usepackage{amsmath,amssymb,amsfonts}%
\usepackage{amsthm}%
\usepackage{mathrsfs}%
\usepackage[title]{appendix}%
\usepackage{xcolor}%
\usepackage{textcomp}%
\usepackage{manyfoot}%
\usepackage{booktabs}%
\usepackage{algorithm}%
\usepackage{algorithmicx}%
\usepackage{algpseudocode}%
\usepackage{listings}%
\usepackage{float}
\usepackage{dcolumn}
\usepackage{bm}
\usepackage{enumitem}
\usepackage{tikz}
\usetikzlibrary{arrows.meta}
\usepackage{subcaption}

\theoremstyle{thmstyleone}%
\theoremstyle{thmstyletwo}%

\theoremstyle{thmstylethree}%

\begin{document}

\title[Article Title]{Non-Radial Free Geodesics. II. In Spatially Curved FLRW Spacetime}


\author[1]{\fnm{Omar} \sur{Nemoul}}\email{omar.nemoul@umc.edu.dz}

\author[1]{\fnm{Hichem} \sur{Guergouri}}\email{hichem.guergouri@univ-bejaia.dz}

\author[1,2]{\fnm{Jamal} \sur{Mimouni}}\email{jamalmimouni@umc.edu.dz}

\affil[1]{\orgdiv{Research Unit in Scientific Mediation}, \orgname{CERIST}, \orgaddress{\city{Constantine}, \postcode{25016}, \country{Algeria}}}

\affil[2]{\orgdiv{Laboratoire de Physique Mathematique et Subatomique (LPMS)}, \orgname{University of Constantine 1}, \orgaddress{\city{Constantine}, \postcode{25017}, \country{Algeria}}}


\abstract{This paper presents an in-depth exploration of timelike free geodesics in spatially curved Friedmann-Lemaître-Robertson-Walker (FLRW) spacetime. A unified approach for these geodesics encompassing both radial and non-radial trajectories across Euclidean, spherical, and hyperbolic geometries is employed. Using the symmetry properties of the system, two constants of motion related to this dynamical system are derived. This treatment facilitates the explicit computation of radial and angular peculiar velocities, along with the evolution of the comoving radial distance and angle over time. The study introduces three distinct methods for characterizing geodesic solutions, further delving into the flatness limit, the null geodesic limit, the radial geodesic limit, the comoving geodesic limit, and the circular orbits. Additionally, we analyze Killing vectors, reflecting the symmetries inherent in the dynamical system. This study provides a more profound understanding of the behavior of free particles as observed from a comoving reference frame.}




\maketitle

\section{Introduction}\label{sec1}
The $\Lambda$CDM cosmological model~\cite{Peebles} has been strongly reinforced by observations of the Cosmic Microwave Background (CMB) from missions like COBE~\cite{COBE}, WMAP~\cite{WMAP}, and Planck~\cite{Planck}, along
with studies highlighting the universe's accelerating expansion rate~\cite{Perlmutter,Riess}. These thorough explorations have consistently revealed the dominance of a mysterious repulsive gravity known as dark energy, driving the universe's accelerating expansion. Additionally, the observable universe exhibits a remarkable degree of spatial homogeneity, isotropy, and geometric flatness on large scales. These characteristics are key to confirming the Friedmann-Lemaître-Robertson-Walker (FLRW) metric as the definitive model for describing our spacetime~\cite{Friedmann,Lemaitre,Robertson,Walker}. The study of geodesics is of crucial importance not only in general relativity~\cite{Einstein} but also in cosmology~\cite{Baumann} , astrophysics, and quantum gravity. A special focus on radially freely-falling test particles in the absence of non-gravitational forces are extensively explored in Refs.~\cite{Whiting,Davis,Gron,Barnes,Kerachian,Vachon,Cotaescu,Omar1}. While several studies~\cite{Whiting,Gron,Barnes,Kerachian,Vachon} have mainly focused on solving geodesic equations, other research efforts~\cite{Cotaescu,Omar1} used the symmetry of the system to construct the exact solution. In our recent study~\cite{Omar2}, we explored the free timelike geodesics for non-radial motion in spatially flat FLRW spacetime, and further applied these insights to the $\Lambda$CDM model.

Building on the results in Ref.~\cite{Omar2}, this paper broadens its scope to include all spaces of maximal symmetry. Mathematically speaking, these spaces are characterized by their constant curvature $k$. It is important to note that the complete connected Riemannian manifolds of a constant curvature can be classified into the Euclidean ($k=0$), spherical ($k>0$), and hyperbolic ($k<0$) geometries, other spaces with the same property are isometric to one of these three classes, by the Killing-Hopf theorem ~\cite{Killing,Hopf}. Accordingly, we shall present a unified formulation for all types of timelike free geodesics, encompassing both radial and non-radial trajectories across different geometric spaces. Initially, we start with a thorough review of the problem in flat space, recalling key findings from Ref.~\cite{Omar2} to lay the groundwork for our extended analysis.

Next, we examine the heart of this study, focusing on the determination of the comoving radial $\chi$ and angular $\phi$ motion for a freely-falling test particle in generally curved FLRW spacetime. It is very challenging to directly solve either the Euler-Lagrange equations~\cite{Euler,Lagrange} or the geodesic equations for the unknown functions $(\chi,\phi)$ and $(t,\chi,\phi)$, respectively. The complexity primarily arises from the intricate nature of the coupled partial differential equations system that emerges when employing both the Euler-Lagrange and geodesic equations method. Fortunately, the inherent symmetry of the physical system offers a key advantage in overcoming this obstacle by yielding two constants of motion used in our approach. The isotropic symmetry of the system concerning the angular variable $\phi$, gives rise to the first constant $B_k$. The derivation of the second constant involves recognizing that the magnitude of peculiar velocity is a 3D scalar vector. By extending its established expression for radial to non-radial motion, we successfully derive the second constant of motion $A_k$, which now accounts for both radial and angular components of the peculiar velocity. In Appendix~\ref{A3}, we rigorously demonstrate that the non-radial solutions satisfy the Euler-Lagrange equations. Subsequently, we establish that the non-radial geodesics in curved space meet several limits: the flatness limit, the radial geodesic limit, the null geodesic limit, and the comoving geodesic limit. Furthermore, we provide simpler alternative expressions for the exact solutions $\chi(t)$ and $\phi(t)$. Our discussion then delves into the Killing vectors associated with the symmetries of the dynamical system with determining their precise expressions.

Throughout this paper, Greek letters $\mu,\nu$, etc., represent spacetime indices, taking values $0$, $1$, $2$, and $3$. Spatial indices, denoted by Latin letters $i,j$, etc., have values $1$, $2$, and $3$. We adopt a metric signature of $(+,-,-,-)$ in a spacetime coordinate system defined by cosmic time $t$ and comoving spatial coordinates $x^i$. $(x,y,z)$ and $(\chi,\theta,\phi)$ represent the comoving Cartesian and spherical coordinate systems, respectively. Moreover, our analysis employs a unit system where the speed of light is normalized to $c=1$.
\section{Non-Radial Geodesics in Spatially Flat FLRW Spacetime}\label{sec2}
Before addressing the non-radial geodesics for free motion in general FLRW spacetime, it is useful to first examine the problem in the flat case. This was recently undertaken in our paper~\cite{Omar2}, where motion is constrained to the $(xy)$ plane by setting $\theta=\frac{\pi}{2}$. This constraint simplifies the spacetime interval to
\begin{align}
ds^2&=dt^2-\gamma_{ij}dx^idx^j\nonumber\\
& = dt^2 - a^2(t)(d\chi^2 + \chi^2 d\phi^2).\label{eq1}
\end{align}
where $\gamma_{ij}$ represents the 3D spatial metric, while $a(t)$ is the expansion scalar factor, a function describing how spatial dimensions scale over time. The action for a freely-falling particle is expressed as
\begin{align}
S[\chi(t),\phi(t)] &= -m\int ds = -m\int dt \sqrt{1 - a^2(t) \dot{\chi}^2 - a^2(t) \chi^2 \dot{\phi}^2}.\label{eq2}
\end{align}
The corresponding Lagrangian is
\begin{align}
L(\chi,\dot{\chi},\dot{\phi},t) = -m\sqrt{1 - a^2(t) \dot{\chi}^2 - a^2(t) \chi^2 \dot{\phi}^2}.\label{eq3}
\end{align}
To ensure that the Lagrangian is expressed in energy units and to obtain the correct non-relativistic limit in accordance with Newton's law, we've included the factor “$-m$”. The squared magnitude of the peculiar velocity is given
by
\begin{align}
v_{\text{pec}}^2(t) &= a^2(t) \gamma_{ij} \dot{x}^i \dot{x}^j\nonumber \\
&= a^2(t) \dot{\chi}^2 + a^2(t) \chi^2 \dot{\phi}^2.\label{eq4}
\end{align}
Subsequently, two symmetries of this Lagrangian given by the Killing vectors
\begin{subequations}
\begin{align}
&\xi_{A} = \pm \sqrt{A^2 - \frac{B^2}{\chi^2}} \partial_{\chi} + \frac{B}{\chi^2} \partial_{\phi},\label{eq5a} \\
&\xi_{B} = \partial_{\phi},\label{eq5b}
\end{align}
\end{subequations}
are used to derive two constants of motion 
\begin{subequations}
\begin{align}
&A = \text{sgn}(A)a^2(t) \sqrt{\frac{ \dot{\chi}^2+ \chi^2 \dot{\phi}^2}{1 - a^2(t) \dot{\chi}^2 - a^2(t) \chi^2 \dot{\phi}^2}}.\label{eq6a} \\
&B = \frac{a^2(t) \chi^2 \dot{\phi}}{\sqrt{1 - a^2(t) \dot{\chi}^2 - a^2(t) \chi^2 \dot{\phi}^2}}.\label{eq6b}
\end{align}
\end{subequations}
It can be seen that the geodesic solution does not depend on $\text{sgn}(A)$, where both “$A$” and “$-A$” yield the same physics. As delineated in Ref.~\cite{Omar2}, the exact solution can be characterized by three distinct cases:
\begin{itemize}[leftmargin=*]
\item \textbf{In terms of $(\chi_i, \phi_i, A, B,\text{sgn}(\dot{\chi}_i))$:}
\end{itemize}
\begin{subequations}
\label{eq7}
\begin{align}
&\chi(t; \chi_i, A, B, \text{sgn}(\dot{\chi}_i)) = \sqrt{\left(\int_{t_i}^{t} \frac{|A|}{a(t') \sqrt{a^2(t') + A^2}} dt' + \text{sgn}(\dot{\chi}_i) \sqrt{\chi_i^2 - \frac{B^2}{A^2}}\right)^2 + \frac{B^2}{A^2}},\label{eq7a} \\
&\phi(t; \chi_i, \phi_i, A, B, \text{sgn}(\dot{\chi}_i)) = \int_{t_i}^{t}  \frac{\frac{B}{\chi^2(t'; \chi_i, A, B, \text{sgn}(\dot{\chi}_i))}}{a(t') \sqrt{a^2(t') + A^2}} dt' + \phi_i,\label{eq7b}
\end{align}
\end{subequations}
with the initial radial distance $\chi_i\geq\left|\frac{B}{A}\right|$, the initial angle $-\pi<\phi_i\leq\pi$, arbitrary real constants of motion $A$ and $B$, where $A=0\Rightarrow B=0$ (See Ref.~\cite{Omar2}) and the sign of the initial comoving radial velocity $\text{sgn}(\dot\chi_i)=\pm$ or $0$.
\begin{itemize}[leftmargin=*]
\item \textbf{In terms of $(\chi_i,\phi_i, v_{\chi, i}, v_{\phi, i})$:}
\end{itemize}
{\small
\begin{subequations}
\label{eq8}
\begin{align}
&\chi(t; \chi_i, v_{\chi, i}, v_{\phi, i}) = \sqrt{\biggl( \int_{t_i}^{t} \frac{a_i}{a(t')} \sqrt{\frac{v_{\chi, i}^2 + v_{\phi, i}^2}{a^2(t')(1 - v_{\chi, i}^2 - v_{\phi, i}^2) + a_i^2(v_{\chi, i}^2 + v_{\phi, i}^2)}} \, dt' +\frac{\chi_i v_{\chi, i}}{\sqrt{v_{\chi, i}^2 + v_{\phi, i}^2}} \biggr)^2+\frac{\chi_i^2 v_{\phi, i}^2}{v_{\chi, i}^2 + v_{\phi, i}^2}}\label{eq8a}\\
&\phi(t; \chi_i, \phi_i,v_{\chi, i}, v_{\phi, i}) = \int_{t_i}^{t} \frac{a_i}{a(t')} \frac{\frac{\chi_i v_{\phi, i}}{\chi^2(t'; \chi_i, v_{\chi, i}, v_{\phi, i})}}{\sqrt{a^2(t')(1 - v_{\chi, i}^2 - v_{\phi, i}^2) + a_i^2(v_{\chi, i}^2 + v_{\phi, i}^2)}} \, dt' + \phi_i,\label{eq8b}
\end{align}
\end{subequations}
}
with $\chi_i\geq0$, $-\pi<\phi_i\leq\pi$, and $0\leq v_{\chi, i}^2 + v_{\phi, i}^2 \leq 1$. Here, $v_{\chi, i}$ and $v_{\phi, i}$ represent the initial radial and angular peculiar velocity components at a specific initial time $t_i$, respectively. The notation $a_i=a(t_i)$ is employed. 
\begin{itemize}[leftmargin=*]
\item \textbf{In terms of $(\chi_i,\phi_i, v_{\text{pec}, i},\psi_i)$:}
\end{itemize}
\begin{subequations}
\label{eq9}
\begin{align}
&\chi(t; \chi_i, v_{\text{pec}, i},\psi_i) = \sqrt{\left( \int_{t_i}^{t} \frac{a_i}{a(t')} \frac{v_{\text{pec}, i}}{\sqrt{a^2(t')(1 - v_{\text{pec}, i}^2) + a_i^2v_{\text{pec}, i}^2}} \, dt' + \chi_i\cos\psi_i \right)^2 + \chi^2_i\sin^2\psi_i}, \label{eq9a}\\
&\phi(t; \chi_i,\phi_i, v_{\text{pec}, i},\psi_i) = \int_{t_i}^{t} \frac{a_i}{a(t')} \frac{\frac{\chi_iv_{\text{pec}, i}\sin{\psi_i}}{\chi^2(t';\chi_i, v_{\text{pec}, i},\psi_i)}}{\sqrt{a^2(t')(1 - v_{\text{pec}, i}^2) + a_i^2v_{\text{pec}, i}^2}} \, dt' + \phi_i,\label{eq9b}
\end{align}
\end{subequations}
with $\chi_i\geq0$, $-\pi<\phi_i\leq\pi$, $0\leq v_{\text{pec}, i}\leq1$, $-\pi<\psi_i\leq\pi$, and $-\pi<\psi_i\leq\pi$. Here, $v_{\text{pec},i}$ represents the magnitude of $\vec v_{\text{pec}, i}$, and $\psi_i$ indicates the deviation angle of $\vec v_{\text{pec},i}$ from the radial axis (deviation angle) at $t = t_i$. They are given by
\begin{subequations}
\begin{align}
v_{\text{pec},i} &= \sqrt{v_{\chi,i}^2 + v_{\phi,i}^2},\label{eq10a}\\
\psi_i &= \arctan\left(\frac{v_{\phi,i}}{v_{\chi,i}}\right).\label{eq10b}
\end{align}
\end{subequations}
Note that in this case of flat space and for given constants of motion $A$ and $B$, the smallest possible distance to which the free particle can approach the origin at $t_*$ is $\chi_* = |B/A|$, corresponding to $\dot{\chi}_* = 0$.
\subsection{Radial-Geodesic Limit}
The geodesic for Radial Motion (RM) can be obtained by setting $B=0$ in Eqs.~\eqref{eq7}, or $v_{\phi,i}$ in Eqs.~\eqref{eq8}, or $\psi_i = 0, \pi$ in Eqs.~\eqref{eq9}. It gives
\begin{subequations}
\label{eq11}
\begin{align}
&\chi(t) = \left| \int_{t_i}^{t} \frac{1}{a(t')} \frac{A}{\sqrt{a^2(t') + A^2}} \, dt' + \chi_i \right|,\label{eq11a}\\
&\phi(t)=\phi_i.\label{eq11b}
\end{align}
\end{subequations}
Here, we use the redundancy of $\text{sgn}(A)$ and choose to align it with $\text{sgn}(\dot\chi_i)$. This solution can be directly derived by starting from the following spacetime interval for radial motion
\begin{equation}
ds^2_{\text{RM}} = dt^2 - a^2(t)d\chi^2.\label{eq12}
\end{equation}
The action for radial motion only depends on $\chi(t)$ as
\begin{equation}
S_{\text{RM}}[\chi(t)] =-m\int dt \sqrt{1 - a^2(t) \dot{\chi}^2}.\label{eq13}
\end{equation}
The corresponding Lagrangian is
\begin{equation}
L_{\text{RM}}(\dot{\chi}(t),t) = -m\sqrt{1 - a^2(t) \dot{\chi}^2}.\label{eq14}
\end{equation}
Notice that this Lagrangian exhibits a transnational symmetry in the radial distance $\chi$, leading to the following conserved quantity
\begin{equation}
A_{\text{RM}}=A\big|_{B=0}=\frac{a^2(t) \dot{\chi}}{\sqrt{1 - a^2(t) \dot{\chi}^2}}.\label{eq15}
\end{equation}
In this case, the squared magnitude of the peculiar velocity $v_{\text{pec}}^2$ is given by
\begin{align}
v_{\text{pec}}^2(t) = v_{\chi}^2(t)
= a^2(t) \dot{\chi}^2
= \frac{A_{\text{RM}}^2}{a^2(t) + A_{\text{RM}}^2}.\label{eq16}
\end{align}
The radial solution~\eqref{eq11} corresponding to the action~\eqref{eq13} has been derived in various works~\cite{Whiting,Davis,Gron,Barnes,Kerachian,Vachon,Cotaescu,Omar1}. It is important to note that this treatment of the radial geodesic does not explicitly depend on the geometry of the 3D space. Both flat, spherical, and hyperbolic spaces yield the same solution~\eqref{eq11} for radial motion. The only distinction among them lies in the expression of the scale factor $a(t)$.
\subsection{Null and Comoving Geodesic Limits}
The null geodesic can be found by having $A, B \to +\infty$ in Eqs.~\eqref{eq7}, or $v_{\chi,i}^2 + v_{\phi,i}^2 = 1$ in Eqs.~\eqref{eq8}, or $v_{pec,i} = 1$ in Eqs.~\eqref{eq9}. On the other hand, the comoving geodesic arises when setting all parameters $A, B, v_{\chi,i}, v_{\phi,i}$, and $v_{pec,i}$ to be zero.\\

In the next section, we shall generalize this analysis to spatially curved FLRW spacetime, not limited to flat space.
\section{Non-Radial Geodesics in generally Curved FLRW Spacetime}\label{sec3}
We now turn to a homogeneous and isotropic 3D space, whose arbitrary spatial geometry is captured by the curvature parameter $k$. This parameter allows for three distinct geometrical spaces:
\begin{equation}
\begin{array}{ll}
 \text{Spherical space}, & k > 0 \\
\text{Flat space}, & k = 0\\
\text{Hyperbolic space}, & k < 0
\end{array}\nonumber
\end{equation}
Here, the curvature parameter $k$ is defined in terms of the curvature radius $\rho$ as follows
\begin{equation}
k = \frac{K}{\rho^2},\label{eq17}
\end{equation}
where $K = \text{sgn}(k)$ denotes the curvature sign, taking values -1, 0, and +1 for hyperbolic, flat, and spherical geometries, respectively. The curvature parameter $k$ can also be expressed in terms of measurable quantities as
\begin{equation}
k = -\frac{H_0^2 \Omega_{k,0}}{c^2},\label{eq18}
\end{equation}
where $H_0$ is the Hubble parameter at the present time, and $c$ is the speed of light. $\Omega_{k,0}$ represents today's curvature density parameter, given by
\begin{equation}
\Omega_{k,0} = 1 - \Omega_{r,0} - \Omega_{m,0} - \Omega_{\Lambda,0}.\label{eq19}
\end{equation}
Here, $\Omega_{r,0}$, $\Omega_{m,0}$, and $\Omega_{\Lambda,0}$ represent the radiation, matter, and cosmological constant density parameters at the present time, respectively.
Now, to address the problem in a generally 3D space, one can simply replace $\chi$ in the flat case with $\frac{1}{\sqrt{k}} \sin(\sqrt{k} \chi)$ for spherical space, and $\frac{1}{\sqrt{|k|}} \sinh(\sqrt{|k|} \chi)$ for the hyperbolic case. Accordingly, let's define the spacetime interval for motion restricted to the $(xy)$ plane as
\begin{align}
ds_k^2&=dt^2-\gamma^{(k)}_{ij}dx^idx^j\nonumber\\
&= dt^2 - a^2(t)(d\chi^2 + S_k^2(\chi)d\phi^2)
,\label{eq20}
\end{align}
where $\gamma_{ij}^{(k)}$ represents the 3D spatial metric for the three classes of maximally symmetric space ($K=0,-1,+1$).
The $S_k$ function is defined as
\begin{align}
S_k(\chi) = 
\begin{cases}
    \frac{1}{\sqrt{k}} \sin(\sqrt{k} \chi), & k > 0 \\
    \chi, & k = 0 \\
    \frac{1}{\sqrt{-k}} \sinh(\sqrt{-k} \chi), & k < 0
\end{cases},\label{eq21}
\end{align}
with its inverse $S_k^{-1}$ function
\begin{align}
S_k^{-1}(\alpha) = 
\begin{cases}
    \frac{1}{\sqrt{k}} \arcsin(\sqrt{k} \alpha), & k > 0 \\
    \alpha, & k = 0 \\
    \frac{1}{\sqrt{-k}} \text{arcsinh}(\sqrt{-k} \alpha), & k < 0
\end{cases}.\label{eq22}
\end{align}
We define the $C_k$ function as
\begin{align}
C_k(\chi) = \frac{dS_k(\chi)}{d\chi} = 
\begin{cases}
    \cos(\sqrt{k} \chi), & k > 0 \\
    1, & k = 0 \\
    \cosh(\sqrt{-k} \chi), & k < 0
\end{cases},\label{eq23}
\end{align}
with its inverse $C_k^{-1}$ function
\begin{align}
C_k^{-1}(\alpha) = 
\begin{cases}
    \frac{1}{\sqrt{k}} \arccos{(\alpha)}, & k > 0 \\
    \frac{1}{\sqrt{-k}} \text{arccosh}( \alpha), & k < 0
\end{cases}.\label{eq24}
\end{align}
and finally, the $T_k$ function
\begin{align}
T_k(\chi) = \frac{S_k(\chi)}{C_k(\chi)} = 
\begin{cases}
    \frac{1}{\sqrt{k}}\tan(\sqrt{k} \chi), & k > 0 \\
    \chi, & k = 0 \\
    \frac{1}{\sqrt{-k}}\tanh(\sqrt{-k} \chi), & k < 0
\end{cases},\label{eq25}
\end{align}
with its inverse $T_k^{-1}$ function
\begin{align}
T_k^{-1}(\alpha) = 
\begin{cases}
    \frac{1}{\sqrt{k}} \text{arcctan}(\sqrt{k}\alpha), & k > 0 \\
    \alpha, & k = 0 \\
    \frac{1}{\sqrt{-k}} \text{arctanh}(\sqrt{-k}\alpha), & k < 0
\end{cases}.\label{eq26}
\end{align}
They satisfy the following identities
\begin{subequations}
\begin{align}
 &C_k^2(\chi) + k S_k^2(\chi) = 1,\label{eq27a} \\
&1 + k T_k^2(\alpha)= \frac{1}{C_k^2(\alpha)}.\label{eq27b}
\end{align}
\end{subequations}
The first identity includes the Pythagorean trigonometric identity for $k > 0$ and the hyperbolic identity for $k < 0$. In this context, the comoving radial distance $\chi$ assumes real, positive values for $k \leq 0$. However, for $k > 0$, $\chi$ is confined to a compact range, specifically $0 \leq \chi \leq \frac{\pi}{\sqrt{k}}$. Additionally, the function $C_k$ is not invertible for $k = 0$, as it does not constitute an injective (one-to-one) mapping in this case.
\subsection{Geodesics From Euler-Lagrange Equations}\label{sec3.1}
Let us now formulate the action for a space generally curved, characterized by the curvature parameter $k$. We have
\begin{align}
S_k[\chi(t), \phi(t)]= -m \int ds_k = -m \int dt \sqrt{1 - a^2(t) \dot{\chi}^2 - a^2(t) S_k^2(\chi) \dot{\phi}^2}.\label{eq28}
\end{align}
In this context, the Lagrangian is expressed as
\begin{equation}
L_k(\chi, \dot{\chi}, \dot{\phi}, t) = -m \sqrt{1 - a^2(t) \dot{\chi}^2 - a^2(t) S_k^2(\chi) \dot{\phi}^2}.\label{eq29}
\end{equation}
The squared magnitude of the peculiar velocity takes the form
\begin{align}
v_{\text{pec}}^2(t) &= a^2(t) \gamma_{ij}^{(k)} \dot{x}^i \dot{x}^j,\label{eq30}
\end{align}
where $\gamma_{ij}^k$ represents the 3D spatial metric. This can be decomposed into radial and angular components, resulting in
\begin{equation}
v_{\text{pec}}^2(t) = v_{\chi}^2(t) + v_{\phi}^2(t),\label{eq31}
\end{equation}
where
\begin{subequations}
\begin{align}
v_{\chi}^2(t) &= a^2(t) \dot{\chi}^2,\label{eq32a} \\
v_{\phi}^2(t) &= a^2(t) S_k^2(\chi) \dot{\phi}^2.\label{eq32b}
\end{align}
\end{subequations}
Following the methodology used in the recent work Ref.~\cite{Omar2}, where the exact solution for non-radial timelike geodesics in the flat case were constructed, we apply the Euler-Lagrange equations for $(\chi(t), \phi(t))$, yielding
\begin{subequations}
\begin{align}
\frac{d}{dt} \left[ \frac{a^2(t) \dot{\chi}(t)}{\sqrt{1 - a^2(t) \dot{\chi}^2(t) - a^2(t) S_k^2(\chi(t)) \dot{\phi}^2(t)}} \right] &= \frac{S_k(\chi(t)) C_k(\chi(t)) a^2(t) \dot{\phi}^2(t)}{\sqrt{1 - a^2(t) \dot{\chi}^2(t) - a^2(t) S_k^2(\chi(t)) \dot{\phi}^2(t)}},\label{eq33a} \\
\frac{d}{dt} \left[ \frac{a^2(t) S_k^2(\chi(t)) \dot{\phi}(t)}{\sqrt{1 - a^2(t) \dot{\chi}^2(t) - a^2(t) S_k^2(\chi(t)) \dot{\phi}^2(t)}} \right] &= 0.\label{eq33b}
\end{align}
\end{subequations}
\subsection{Determining the Constants of Motion}\label{sec3.2}
It is evident that the Lagrangian exhibits symmetry under any time-independent rotation $\delta$ of the angular variable $\phi$, as it is independent of $\phi$
\begin{equation}
\phi \to \phi' = \phi + \delta.\label{eq34}
\end{equation}
From Noether’s theorem~\cite{Noether}, this symmetry in the dynamical system implies the existence of a conservation law, signified by a conserved quantity that remains invariant over time (constant of motion). From the Euler-Lagrange equation, we derive the constant of motion as
\begin{equation}
\frac{a^2(t) S_k^2(\chi) \dot{\phi}}{\sqrt{1 - a^2(t) \dot{\chi}^2 - a^2(t) S_k^2(\chi) \dot{\phi}^2}} = B_k,\label{eq35}
\end{equation}
where $B_k$ is an arbitrary real number, related to both the initial angular and radial velocities of the particle. It’s important to note that:
\begin{itemize}[leftmargin=*]
    \item $\text{sgn}(\dot{\phi}) = \text{sgn}(B_k)$,
    \item $B_k > 0 \Leftrightarrow \phi(t)$ is an increasing function,
    \item $B_k < 0 \Leftrightarrow \phi(t)$ is a decreasing function,
    \item Radial Geodesics: $B_k = 0 \Leftrightarrow \phi(t)$ is a constant function,
    \item $B_0 = B$, which corresponds to the same constant of motion as in the flat case.
\end{itemize}
By inverting Eq.~\eqref{eq35}, we can obtain the square of the angular peculiar velocity $v_{\phi}^2$ as
\begin{equation}
v_{\phi}^2(t) = a^2(t) S_k^2(\chi) \dot{\phi}^2 = \frac{B_k^2 (1 - a^2(t) \dot{\chi}^2)}{a^2(t) S_k^2(\chi) + B_k^2}.\label{eq36}
\end{equation}
To identify the second constant of motion, we use the fact that the magnitude of the peculiar velocity $v_{\text{pec}}(t)$ is a 3D scalar, invariant under spatial coordinate transformations, as evident from its spatial covariant form in Eq.~\eqref{eq30}. In our homogeneous and isotropic 3D space, any general geodesic curve can be transformed to have a purely radial spatial part through appropriate local spatial coordinates. This allows us to generalize the 3D scalar peculiar velocity expression~\eqref{eq16} for radial motion to include all general geodesic motions. Hence, we have
\begin{align}
v_{\text{pec}}^2(t) &= v_{\chi}^2(t) + v_{\phi}^2(t) \nonumber\\
&= a^2(t) \dot{\chi}^2 + a^2(t) S_k^2(\chi) \dot{\phi}^2 \nonumber\\
&= \frac{A_k^2}{a^2(t) + A_k^2}.\label{eq37}
\end{align}
The real number $A_k$ is now related to both the initial conditions for both radial and angular peculiar velocities. It is worth mentioning that:
\begin{itemize}[leftmargin=*]
    \item $\text{sgn}(A_k)$ is redundant for our physical motion, i.e., does not affect the physical nature of the motion where both “$A_k$” and “$-A_k$” represent the same physics.
    \item Comoving Geodesics: $A_k = 0 \Leftrightarrow v_{\text{pec}}(t) = 0 \Leftrightarrow v_{\chi}(t) = 0$ and $v_{\phi}(t) = 0$
    \item $A_k = 0 \Rightarrow B_k = 0$, although the converse does not necessarily hold true.
    \item Null Geodesics: $v_{\text{pec}}(t) = 1 \Leftrightarrow A_k = +\infty \Leftrightarrow B_k = +\infty$
    \item $A_0 = A$, gives the same constant of motion for the flat case.
\end{itemize}
We have obtained a system of two equations, involving the comoving velocity variables $(\dot{\chi}, \dot{\phi})$. Solving these two equations for $(\dot{\chi}, \dot{\phi})$ gives
\begin{subequations}
\label{eq38}
\begin{align}
&\dot{\chi}(t) = \frac{\text{sgn}(\dot{\chi}(t))}{a(t)} \sqrt{\frac{A_k^2 - \frac{B_k^2}{S_k^2(\chi(t))}}{a^2(t) + A_k^2}},\label{eq38a} \\
&\dot{\phi}(t) =  \frac{\frac{B_k}{S_k^2(\chi(t))}}{a(t) \sqrt{a^2(t) + A_k^2}},\label{eq38b}
\end{align}
\end{subequations}
and by corresponding the peculiar velocity components $(v_{\chi}, v_{\phi})$ are
\begin{subequations}
\label{eq39}
\begin{align}
v_{\chi}(t) &= a(t) \dot{\chi}(t) = \text{sgn}(v_{\chi}(t)) \sqrt{\frac{A_k^2 - \frac{B_k^2}{S_k^2(\chi(t))}}{a^2(t) + A_k^2}},\label{eq39a} \\
v_{\phi}(t) &= a(t) S_k(\chi) \dot{\phi}(t) =  \frac{\frac{B_k}{S_k(\chi(t))}}{\sqrt{a^2(t) + A_k^2}}.\label{eq39b}
\end{align}
\end{subequations}
We note that both radial and angular peculiar velocities decrease over time, approaching zero as the scale factor tends towards infinity.
\subsection{The initial Conditions for Peculiar Velocity Components}\label{sec3.3}
It is clear from Eqs.~\eqref{eq39a} and~\eqref{eq39b} that in order to determine the peculiar velocity components $v_{\chi}(t)$ and $v_{\phi}(t)$ of the particle, the constants of motion $A_k$ and $B_k$ must be fixed. Now, we will express $A_k$ and $B_k$ in terms of the initial radial $v_{\chi,i} = v_{\chi}(t_i)$ and angular $v_{\phi,i} = v_{\phi}(t_i)$ peculiar velocities at a chosen initial time $t_i$. By substituting $t = t_i$ into Eqs.~\eqref{eq39a} and~\eqref{eq39b} and then inverting these equations for $A_k$ and $B_k$, we obtain
\begin{subequations}
\begin{align}
A_k &= \text{sgn}(A) a_i \sqrt{\frac{v_{\chi,i}^2 + v_{\phi,i}^2}{1 - v_{\chi,i}^2 - v_{\phi,i}^2}},\label{eq40a} \\
B_k &= \frac{a_i S_k(\chi_i) v_{\phi,i}}{\sqrt{1 - v_{\chi,i}^2 - v_{\phi,i}^2}}.\label{eq40b}
\end{align}
\end{subequations}
Substituting these derived expressions for $A_k$ and $B_k$ back into Eqs.~\eqref{eq38} and~\eqref{eq39}, the equations for $\dot{\chi}(t)$ and $\dot{\phi}(t)$ are
\begin{subequations}
\begin{align}
\dot{\chi}(t) &= \text{sgn}(\dot{\chi}(t)) \frac{a_i}{a(t)} \sqrt{\frac{v_{\chi,i}^2 + v_{\phi,i}^2 - \frac{S_k^2(\chi_i) v_{\phi,i}^2}{S_k^2(\chi(t))}}{a^2(t)(1 - v_{\chi,i}^2 - v_{\phi,i}^2) + a_i^2 (v_{\chi,i}^2 + v_{\phi,i}^2)}},\label{eq41a} \\
\dot{\phi}(t) &= \frac{a_i}{a(t)}  \frac{\frac{S_k(\chi_i) v_{\phi,i}}{S_k^2(\chi(t))}}{\sqrt{a^2(t)(1 - v_{\chi,i}^2 - v_{\phi,i}^2) + a_i^2 (v_{\chi,i}^2 + v_{\phi,i}^2)}},\label{eq41b}
\end{align}
\end{subequations}
and their corresponding peculiar components
\begin{subequations}
\begin{align}
v_{\chi}(t) &= \text{sgn}(v_{\chi}(t)) \frac{a_i}{a(t)} \sqrt{\frac{v_{\chi,i}^2 + v_{\phi,i}^2 - \frac{S_k^2(\chi_i) v_{\phi,i}^2}{S_k^2(\chi(t))}}{a^2(t)(1 - v_{\chi,i}^2 - v_{\phi,i}^2) + a_i^2 (v_{\chi,i}^2 + v_{\phi,i}^2)}},\label{eq42a} \\
v_{\phi}(t) &=  \frac{\frac{a_i S_k(\chi_i) v_{\phi,i}}{S_k(\chi(t))}}{\sqrt{a^2(t)(1 - v_{\chi,i}^2 - v_{\phi,i}^2) + a_i^2 (v_{\chi,i}^2 + v_{\phi,i}^2)}},\label{eq42b}
\end{align}
\end{subequations}
where $0 \leq v_{\chi,i}^2 + v_{\phi,i}^2 \leq 1$. These expressions provide the comoving velocity components $(\dot{\chi}, \dot{\phi})$, and their corresponding peculiar velocities $(v_{\chi}, v_{\phi})$ at any time $t$, in terms of the initial peculiar velocities $(v_{\chi,i}, v_{\phi,i})$ at an initial time $t_i$. Additionally, using polar coordinates $(v_{\text{pec},i}, \psi_i)$ instead of $(v_{\chi,i}, v_{\phi,i})$ can parameterize the comoving and peculiar velocities more effectively. This polar coordinate system is usseful for determining the comoving and peculiar velocity components as
\begin{subequations}
\begin{align}
\dot{\chi}(t) &= \text{sgn}(\dot{\chi}(t)) \frac{a_i v_{\text{pec},i}}{a(t)} \sqrt{\frac{1 - \frac{S_k^2(\chi_i)}{S_k^2(\chi(t))} \sin^2\psi_i}{a^2(t)(1 - v_{\text{pec},i}^2) + a_i^2 v_{\text{pec},i}^2}},\label{eq43a}\\
\dot{\phi}(t) &= \frac{a_i v_{\text{pec},i}}{a(t)}  \frac{\frac{S_k(\chi_i)}{S_k^2(\chi(t))} \sin\psi_i}{\sqrt{a^2(t)(1 - v_{\text{pec},i}^2) + a_i^2 v_{\text{pec},i}^2}},\label{eq43b}
\end{align}
\end{subequations}
and
\begin{subequations}
\begin{align}
v_{\chi}(t) &= \text{sgn}(v_{\chi}(t)) a_i v_{\text{pec},i} \sqrt{\frac{1 - \frac{S_k^2(\chi_i)}{S_k^2(\chi(t))} \sin^2\psi_i}{a^2(t)(1 - v_{\text{pec},i}^2) + a_i^2 v_{\text{pec},i}^2}},\label{eq44a}\\
v_{\phi}(t) &=  \frac{\frac{S_k(\chi_i)}{S_k(\chi(t))} a_i v_{\text{pec},i} \sin\psi_i}{\sqrt{a^2(t)(1 - v_{\text{pec},i}^2) + a_i^2 v_{\text{pec},i}^2}}.\label{eq44b}
\end{align}
\end{subequations}
Note that at $t = t_i$, inward radial motion corresponds to $\psi_i = 0$, whereas outward radial motion relates to $\psi_i = \pi$.
\subsection{Geodesics Parametrization for General Motion}\label{sec3.4}
Geodesic motion in curved spacetime can be found by integrating Eqs.~\eqref{eq38a} and~\eqref{eq38b}. However, the treatment for the radial $\chi$ equation is non-trivial, this complexity is thoroughly explored in Appendix~\ref{A1}, where a detailed analysis and careful consideration are provided. Now, building upon previous work in this field, we establish three distinct methods to characterize a specific geodesic solution for a freely-falling test particle under any real value of the curvature parameter $k$, as described below:
\begin{itemize}[leftmargin=*]
\item $(\chi_i, \phi_i, A_k, B_k, \text{sgn}(\dot{\chi}_i))$ \textbf{initial conditions:}
\end{itemize}
 Using Eqs.~\eqref{eq38a} and~\eqref{eq38b}, we express the comoving radial distance $\chi$ (as detailed in Appendix~\ref{A1}) and the angle $\phi$ for a free test-particle at any given time $t$, under four specific initial conditions $(\chi_i, \phi_i, A_k, B_k)$ and $\text{sgn}(\dot{\chi}_i)$ as
\begin{subequations}
\label{eq45}
\begin{align}
&\chi(t; \chi_i, A_k, B_k, \text{sgn}(\dot{\chi}_i)) = S_k^{-1}\left(\sqrt{\left(1 - k \frac{B_k^2}{A_k^2}\right) S_k^2(R_k(t)) + \frac{B_k^2}{A_k^2}}\right),\label{eq45a}\\
&\phi(t; \chi_i, \phi_i, A_k, B_k, \text{sgn}(\dot{\chi}_i)) = \int_{t_i}^{t}  \frac{\frac{B_k}{S_k^2(\chi(t'; \chi_i, A_k, B_k, \text{sgn}(\dot{\chi}_i)))}}{a(t') \sqrt{a^2(t') + A_k^2}} dt' + \phi_i.\label{eq45b}
\end{align}
\end{subequations}
Here, the radial function $R_k$ is given in terms of $(\chi_i, A_k, B_k, \text{sgn}(\dot{\chi}_i))$ as
\begin{align}
R_k(t; A_k, B_k, \text{sgn}(\dot{\chi}_i)) &= \int_{t_i}^{t} \frac{|A_k| dt'}{a(t') \sqrt{a^2(t') + A_k^2}} dt' + \text{sgn}(\dot{\chi}_i) T_k^{-1}\left(\sqrt{\frac{S_k^2(\chi_i) - \frac{B_k^2}{A_k^2}}{C_k^2(\chi_i)}}\right),\label{eq46}
\end{align}
with $\text{sgn}(\dot{\chi}_i) = \pm$ or $0$, $A_k = 0 \Rightarrow B_k = 0$. By fixing $A_k$ and $B_k$, the radial solution $\chi(t)$ is constrained according to Eqs.~\eqref{eq45a} and~\eqref{eq46} for different curvature parameters $k$ as follows
\begin{align}
\begin{array}{ll}
 \chi(t) \geq S_k^{-1}\left(\left|\frac{B_k}{A_k}\right|\right), &  k \leq 0  \\
S^{-1}_k\left(\left|\frac{B_k}{A_k}\right|\right) \leq \chi(t) \leq \frac{\pi}{\sqrt{k}} - S^{-1}_k\left( \left|\frac{B_k}{A_k}\right|\right), & k > 0
\end{array}.\label{eq47}
\end{align}
In this context, the initial comoving radial distance $\chi_i$ is constrained as $\chi(t)$ by Eq.~\eqref{eq47}. Three important remarks on Eqs.~\eqref{eq46} and~\eqref{eq47} are worth highlighting:
\begin{enumerate}[leftmargin=*]
    \item For all geometrical cases, the trajectory for general motion has a lower bound defined by $S_k^{-1}(|B_k/A_k|)$, which represents the minimum comoving distance that a freely-falling particle can approach the origin.
    \item The upper bound of these trajectories depends on the specific geometry:
       \begin{enumerate}[leftmargin=*]
            \item Flat and Hyperbolic Geometries: In flat ($k=0$) and hyperbolic ($k<0$) geometries, the general geodesics are unbounded. There is no upper limit to the comoving distance $\chi(t)$ a free particle can travel in these spaces.
            \item Spherical Geometry: Conversely, in spherical geometry ($k>0$), the upper bound of $\chi(t)$ is clearly defined. It is specified by $\pi/\sqrt{k} -  S_k^{-1}(|B_k/A_k|)$. This expression represents the maximum distance a free traveler can reach in spherical space. Additionally, for radial motion ($B_k=0$), the upper bound simplifies to $\pi/\sqrt{k}$. This indicates that the free particle can move to the antipodal point of the sphere.
        \end{enumerate}
    \item The function $R_k$ is termed the `radial function' as it represents the radial distance of the freely-falling particle along a radial geodesic, measured by the closest comoving point in the trajectory to the origin, i.e., the projected point of the origin onto the trajectory, which is $S_k^{-1}(|B_k/A_k|)$ away from the origin (for more detail see Sec. 4 in Ref.~\cite{Omar2}).
\end{enumerate}
\begin{itemize}[leftmargin=*]
\item $(\chi_i, \phi_i, v_{\chi,i}, v_{\phi,i})$ \textbf{initial conditions:}
\end{itemize}
Using Eqs.~\eqref{eq41a} and~\eqref{eq41b}, the radial comoving distance $\chi$ and the angle $\phi$ for a free test-particle at any given time $t$, under initial conditions $(\chi_i, \phi_i, v_{\chi,i}, v_{\phi,i})$, are expressed as
\begin{subequations}
\label{eq48}
\begin{align}
&\chi(t; \chi_i, v_{\chi,i}, v_{\phi,i}) = S_k^{-1}\left(\sqrt{\left(1 - k \frac{S_k^2(\chi_i) v_{\phi,i}^2}{v_{\chi,i}^2 + v_{\phi,i}^2}\right) S_k^2(R_k(t)) + \frac{S_k^2(\chi_i) v_{\phi,i}^2}{v_{\chi,i}^2 + v_{\phi,i}^2}}\right),\label{eq48a}\\
&\phi(t; \chi_i, \phi_i, v_{\chi,i}, v_{\phi,i}) = \int_{t_i}^{t} \frac{a_i}{a(t')}  \frac{\frac{S_k(\chi_i) v_{\phi,i}}{S_k^2(\chi(t'; \chi_i, v_{\chi,i}, v_{\phi,i}))}}{\sqrt{a^2(t')(1 - v_{\chi,i}^2 - v_{\phi,i}^2) + a_i^2 (v_{\chi,i}^2 + v_{\phi,i}^2)}} dt' + \phi_i.\label{eq48b}
\end{align}
\end{subequations}
The radial function $R_k$ in terms of $(\chi_i, v_{\chi,i}, v_{\phi,i})$, takes the following form
\begin{align}
R_k(t; \chi_i, v_{\chi,i}, v_{\phi,i}) &= \int_{t_i}^{t} \frac{a_i}{a(t')} \sqrt{\frac{v_{\chi,i}^2 + v_{\phi,i}^2}{a^2(t')(1 - v_{\chi,i}^2 - v_{\phi,i}^2) + a_i^2 (v_{\chi,i}^2 + v_{\phi,i}^2)}} dt' + T_k^{-1}\left(\frac{v_{\chi,i} |T_k(\chi_i)|}{\sqrt{v_{\chi,i}^2 + v_{\phi,i}^2}}\right).\label{eq49}
\end{align}
where $\chi_i\geq0$, $-\pi < \phi_i \leq \pi$, and $0 \leq v_{\chi,i}^2 + v_{\phi,i}^2 \leq 1$.
\begin{itemize}[leftmargin=*]
\item $(\chi_i, \phi_i, v_{\text{pec},i}, \psi_i)$ \textbf{initial conditions:}
\end{itemize}
For the initial conditions $(\chi_i, \phi_i, v_{\text{pec},i}, \psi_i)$, we can use Eqs.~\eqref{eq43a} and~\eqref{eq43b} to express $\chi$ and $\phi$ of a free test-particle at any given time $t$ as
\begin{subequations}
\label{eq50}
\begin{align}
&\chi(t; \chi_i, v_{\text{pec},i}, \psi_i) = S_k^{-1}\left(\sqrt{\left(1 - k S_k^2(\chi_i) \sin^2\psi_i\right) S_k^2(R_k(t)) + S_k^2(\chi_i) \sin^2\psi_i}\right),\label{eq50a}\\
&\phi(t; \chi_i, \phi_i, v_{\text{pec},i}, \psi_i) = \int_{t_i}^{t} \frac{a_i}{a(t')} \frac{\frac{S_k(\chi_i) v_{\text{pec},i} \sin\psi_i}{S_k^2(\chi(t'; \chi_i, v_{\text{pec},i}, \psi_i))}}{\sqrt{a^2(t')(1 - v_{\text{pec},i}^2) + a_i^2 v_{\text{pec},i}^2}} dt' + \phi_i,\label{eq50b}
\end{align}
\end{subequations}
where the radial function $R_k$ is given in terms of $(\chi_i, v_{\text{pec},i}, \psi_i)$ as
\begin{align}
R_k(t; \chi_i, v_{\text{pec},i}, \psi_i) &= \int_{t_i}^{t} \frac{a_i}{a(t')}  \frac{v_{\text{pec},i}}{\sqrt{a^2(t')(1 - v_{\text{pec},i}^2) + a_i^2 v_{\text{pec},i}^2}} dt' + T_k^{-1}\left(|T_k(\chi_i)| \cos\psi_i\right),\label{eq51}
\end{align}
where $\chi_i\geq0$, $-\pi < \phi_i \leq \pi$, $0 \leq v_{\text{pec},i} \leq 1$, and $-\pi < \psi_i \leq \pi$.
\\

In the first parametrization~\eqref{eq45}, when $A_k$ and $B_k$ are fixed, the initial comoving radial distance $\chi_i$ must be constrained within a specific range given by Eq.~\eqref{eq47}. However, one can show that in the second~\eqref{eq48} and third~\eqref{eq50} parametrizations, the initial comoving distance $\chi_i$ is independent of other initial parameters. Therefore, using these latter two formulations is more practical for describing non-radial solutions.
\subsection{The Flat Geodesic Limit}\label{sec3.5}
Setting $k=0$ in Eqs.~\eqref{eq45},~\eqref{eq48}, and~\eqref{eq50}, reduces the expressions to the exact solutions found in equations~\eqref{eq7},~\eqref{eq8}, and~\eqref{eq9}, respectively, corresponding to the non-radial geodesic solution in flat 3D space. This demonstrates the consistency of the general formulation with the well-known results in Euclidean geometry.
\subsection{The Null Geodesic Limit}\label{sec3.6}
To explore the null geodesic limit, we use the parametrization $(\chi_i,\phi_i,v_{\text{pec},i},\psi_i)$ and set $v_{\text{pec},i}=1$. This leads to the following solutions
\begin{subequations}
\begin{align}
&\chi(t; \chi_i, 1, \psi_i) = S_k^{-1}\left(\sqrt{\left(1 - k S_k^2(\chi_i) \sin^2\psi_i\right) S_k^2(R_k(t)) + S_k^2(\chi_i) \sin^2\psi_i}\right),\label{eq52a}\\
&\phi(t; \chi_i, \phi_i, 1, \psi_i) = \int_{t_i}^{t} \frac{1}{a(t')} \frac{S_k(\chi_i) \sin\psi_i}{S_k^2(\chi(t'; \chi_i, 1, \psi_i))} dt' + \phi_i.\label{eq52b}
\end{align}
\end{subequations}
The radial function $R_k$ for this limit is given as:
\begin{equation}
R_k(t) = \int_{t_i}^{t} \frac{dt'}{a(t')} + T_k^{-1}\left(|T_k(\chi_i)| \cos\psi_i\right).\label{eq53}
\end{equation}
\subsection{The Radial Geodesic Limit}\label{sec3.7}
To apply the radial geodesic limit, we use the parametrization $(\chi_i, \phi_i, A_k, B_k)$ and set $B_k = 0$. The comoving solution thus obtained is
\begin{subequations}
\begin{align}
 &\chi(t; \chi_i, A_k, 0, \text{sgn}(\dot{\chi}_i))= |R_k(t)|,\label{eq54a}\\
&\phi(t; \chi_i, \phi_i, A_k, 0, \text{sgn}(\dot{\chi}_i)) = \phi_i.\label{eq54b}
\end{align}
\end{subequations}
The radial function $R_k$ in this context is expressed as:
\begin{equation}
R_k(t) = \int_{t_i}^{t} \frac{A_k dt'}{a(t') \sqrt{a^2(t') + A_k^2}} dt' + \chi_i.\label{eq55}
\end{equation}
where we set $\text{sgn}(A_k)=\text{sgn}(\dot{\chi}_i)$. These Eqs.~\eqref{eq54a} and~\eqref{eq54b} are identical to the solution~\eqref{eq11a} and~\eqref{eq11b} for radial motion. Noting that the spacetime interval for radial motion in Eq.~\eqref{eq12} has the same form across flat, spherical, and hyperbolic geometries. It is unsurprising that the solution in the radial limit does not explicitly depend on the curvature parameter $k$. However, the dependence on geometry in this case is implicitly manifested through the scale factor $a(t)$.
\subsection{Circular Orbits}\label{sec3.8}
These are orbits with a constant radius over time, indicated by $\chi(t) = \chi_i$, resulting in a radial peculiar velocity of $v_{\chi}(t) = v_{\chi,i} = 0$. It can be demonstrated, using Eqs~\eqref{eq42a} and~\eqref{eq48a}, that circular orbits must fulfill the following condition: $S_k^2 (\chi_i ) = \frac{1}{k}$. As expected, this condition uniquely holds for spherical geometry where $k > 0$, leading to $\chi(t) = \chi_i = \frac{\pi}{2\sqrt{k}}$, describing a great circle trajectory that remains equidistant from the origin.
\section{Alternative Expression for the Comoving Radial Distance}\label{sec4}
Applying identity~\eqref{eq27a} in Eq.~\eqref{eq45a}, the comoving radial distance can be simplified using the $C_k$ function for cases where $k \neq 0$. The expression for $\chi(t)$ is:
\begin{equation}
\chi(t) = C_k^{-1}\left(\sqrt{1 - k \frac{B_k^2}{A_k^2}} C_k(R_k(t))\right).\label{eq56}
\end{equation}
Here, the radial function $R_k$ is provided in Eq.~\eqref{eq46}, $C_k$ and its inverse $C^{-1}_k$ is defined in Eqs.~\eqref{eq23} and~\eqref{eq24}. In terms of $(\chi_i,\phi_i,v_{\chi,i},v_{\phi,i})$, the comoving radial distance is:
\begin{equation}
\chi(t) = C_k^{-1}\left(\sqrt{1 - k \frac{S_k^2(\chi_i) v_{\phi,i}^2}{v_{\chi,i}^2 + v_{\phi,i}^2}} C_k(R_k(t))\right),\label{eq57}
\end{equation}
where $R_k$ is now specified by Eq.~\eqref{eq49}. Similarly, for initial conditions $(\chi_i,\phi_i,v_{\text{pec},i},\psi_i)$, $\chi(t)$ is expressed as:
\begin{equation}
\chi(t) = C_k^{-1}\left(\sqrt{1 - k S_k^2(\chi_i) \sin^2\psi_i} C_k(R_k(t))\right),\label{eq58}
\end{equation}
where $R_k$ is outlined in Eq.~\eqref{eq51}. In this case where the exact solution is undefined for $k=0$, we cannot directly set $k=0$. Instead, we should consider the limit where $\chi,\chi_i \ll \frac{1}{\sqrt{k}}$, or take the direct limit as $k \to 0$. One can check that when this limit is applied in Eqs.~\eqref{eq56},~\eqref{eq57}, and~\eqref{eq58}, we obtain the correct solutions for the flat 3D space as outlined in Eqs.~\eqref{eq7},~\eqref{eq8}, and~\eqref{eq9}, respectively.
\section{Alternative Expression for the Angle}\label{sec5}
It is clear that free geodesics in the comoving frame correspond to straight-line geodesics in flat geometry, great circles in spherical geometry, and hyperbolic lines in hyperbolic geometry. As a result, $\phi(t)$ is explicitly dependent only on the radial distance $\chi(t)$ and not on the scale factor $a(t)$. Accordingly, we can derive the geometric formula for the trajectory by eliminating $a(t)$, which has been accomplished in Appendix~\ref{A2}, where an alternative formula describing the real trajectory of the freely-falling particle in flat, spherical, and hyperbolic geometries is presented. The trajectory can be expressed in terms of $\phi(t)$ as
\begin{flalign}
\phi(t) = \sin(B_k) \left[\text{sgn}(\dot{\chi}(t)) \arctan\left(\sqrt{\frac{A_k^2 S_k^2(\chi(t)) - B_k^2}{B_k^2 C_k^2(\chi(t))}}\right) - \text{sgn}(\dot{\chi}_i) \arctan\left(\sqrt{\frac{A_k^2 S_k^2(\chi_i) - B_k^2}{B_k^2 C_k^2(\chi_i)}}\right)\right] + \phi_i.\label{eq59}&&
\end{flalign}
In terms of $(\chi_i, \phi_i, v_{\chi,i}, v_{\phi,i})$
{\small
\begin{equation}
\phi(t) = \sin(v_{\phi,i}) \left[ \text{sgn}(\dot{\chi}(t)) \arctan\left(\sqrt{\frac{S_k^2(\chi(t)) (v_{\chi,i}^2 + v_{\phi,i}^2) - S_k^2(\chi_i) v_{\phi,i}^2}{C_k^2(\chi(t)) S_k^2(\chi_i) v_{\phi,i}^2}}\right) - \text{sgn}(v_{\chi,i}) \arctan\left( \frac{\left|\frac{v_{\chi,i}}{v_{\phi,i}}\right|}{\left|C_k(\chi_i)\right|} \right) \right] + \phi_i.\label{eq60}
\end{equation}}
In terms of $(\chi_i, \phi_i, v_{\text{pec},i}, \psi_i)$
{\small
\begin{equation}
\phi(t) = \sin(\psi_i(\pi - \psi_i)) \left[ \text{sgn}(\dot{\chi}(t)) \arctan\left(\sqrt{\frac{S_k^2(\chi(t)) - S_k^2(\chi_i) \sin^2\psi_i}{C_k^2(\chi(t)) S_k^2(\chi_i) \sin^2\psi_i}}\right) - \text{sgn}\left(\frac{\pi}{2} - |\psi_i|\right) \arctan\left(\frac{\left|\cot\psi_i\right|}{\left|C_k(\chi_i)\right|} \right) \right] + \phi_i.\label{eq61}
\end{equation}}
where we have used the sign relations:
\begin{subequations}
\begin{align}
\text{sgn}(\dot{\chi}_i) &= \text{sgn}(v_{\chi,i}) = \text{sgn}\left(\frac{\pi}{2} - |\psi_i|\right),\label{eq62a}\\
\text{sgn}(B) &= \text{sgn}(v_{\phi,i}) = \text{sgn}(\psi_i(\pi - \psi_i)).\label{eq62}
\end{align}
\end{subequations}
\section{Symmetry and Killing vectors}\label{sec6}
Building on the analysis presented in subsection 4.I. of our recent work~\cite{Omar2}, we deduce the Killing vectors $\xi_{A_k}$ and $\xi_{B_k}$ associated with the constants of motion $A_k$ and $B_k$, respectively, as
\begin{subequations}
\begin{align}
&\xi_{A_k} = \pm\sqrt{A_k^2 - \frac{B_k^2}{S_k^2(\chi)}} \partial_\chi + \frac{B_k}{S_k^2(\chi)} \partial_\phi,\label{eq63a}\\
&\xi_{B_k} = \partial_\phi.\label{eq63b}
\end{align}
\end{subequations}
Four important points are to be highlighted in the context of Killing vectors and their relation to different geometrical spaces:
\begin{enumerate}[leftmargin=*]
    \item For the flat case $k=0$, the constants of motion simplify to $A_0 = A$, and the associated Killing vector $\xi_{A_0} = \xi_A$, where $\xi_A$ is expressed in Eq.~\eqref{eq5a} for the flat case discussion.
    \item The Killing vector $\xi_{B_k}$ related to rotational symmetry, maintains the same form across all geometries: flat, spherical, and hyperbolic. This universality underscores the fundamental nature of the isotropic symmetry in various curved spaces.
     \item It can be easily shown that the transformation generated by these Killing vectors~\eqref{eq63a} and~\eqref{eq63b} render the Lagrangian~\eqref{eq29} invariant.
    \item The derived Killing vectors $\xi_{A_k}$ and $\xi_{B_k}$ satisfy the Killing equation, representing the isometry condition for distance-preserving transformations between metric spaces
    \begin{equation}
    \nabla_{\mu} \xi_{\nu} + \nabla_{\nu} \xi_{\mu} = 0
    \end{equation}
    Here, $\nabla_{\mu}$ is the covariant derivative, defined by the Christoffel symbols for the spacetime interval.
\end{enumerate}
\section{Conclusion}\label{sec7}
In conclusion, this analysis serves as a direct extension of our prior work presented in~\cite{Omar2}. While our earlier study explored non-radial timelike geodesics within a spatially flat FLRW spacetime, this work has expanded it to encompass all maximally symmetric geometries: flat, spherical, and hyperbolic. This broader analysis provides a unified formulation, elegantly expressed in terms of the curvature parameter $k$. The approach we followed is based on using the symmetry of the system to construct two conserved quantities $A_k$ and $B_k$, which makes for a straightforward and simpler determination of the radial $\dot{\chi}$ and the angular $\dot{\phi}$ velocities, along with the evolution of the comoving radial distance $\chi$ and the angle $\phi$. This offers an advantageous alternative to traditional Euler-Lagrange or even the geodesic equation methods. In this framework, geodesics are fully determined by three appropriately parameterized initial conditions $(\chi_i,\phi_i, A_k,B_k)$ as detailed in Eqs.~\eqref{eq45}, $(\chi_i,\phi_i, v_{\chi,i},v_{\phi,i})$ in Eqs.~\eqref{eq48}, and $(\chi_i,\phi_i, v_{\text{pec},i},\psi_i)$ in Eqs.~\eqref{eq50}. Furthermore, we have elucidated the flatness limit, the radial geodesic limit, the null geodesic limit, and the comoving geodesic limit for the established non-radial solution. Our study of general geodesic motion has produced alternative and simpler expressions for the comoving radial solution $\chi(t)$, specifically in Eqs.~\eqref{eq56},~\eqref{eq57}, and~\eqref{eq58}. However, these expressions are only valid for non-flat spatial geometries ($k\neq0$). Moreover, we have derived alternative geometrical formulations for the angle $\phi(t)$, presented in Eqs.~\eqref{eq59},~\eqref{eq60}, and~\eqref{eq61}, which are only dependent on the comoving radial distance $\chi(t)$. Lastly, we have identified the Killing vectors~\eqref{eq63a} and~\eqref{eq63b} corresponding to the conserved quantities $A_k$ and $B_k$. Our paper provides deeper insights into the behavior of free particles from our perspective, encompassing not only radial but also non-radial motions across various curved spacetime geometries. We intend to extend this investigation to incorporate the gravitational influence of galaxies and Earth's peculiar velocity relative to the CMB.
\appendix
\section{Proving the Exact Solution for the Comoving Radial Distance}\label{A1}
The comoving radial distance for free non-radial geodesic motion in a generally FLRW spacetime can be determined by integrating Eq.~\eqref{eq38a}.This process involves separating the coordinate variable $\chi$ from the time coordinate $t$. Once separated, integration yields
\begin{equation}
\int_{\chi_i}^{\chi} \frac{d\chi'}{\sqrt{A_k^2 - \frac{B_k^2}{S_k^2(\chi')}}} = \int_{t_i}^{t} \frac{\text{sgn}(\dot{\chi}(t')) dt'}{a(t') \sqrt{a^2(t') + A_k^2}}.\label{eqA1}
\end{equation}
The left-hand side of this equation can be integrated for each geometric space using the following formulae
\begin{subequations}
\begin{align}
\text{for } k>0: & \int  \frac{d\chi}{\sqrt{A_k^2 - \frac{kB_k^2}{\sin^2(\sqrt{k} \chi)}}} = \frac{1}{|A_k| \sqrt{k}} \arctan\left(\sqrt{\frac{\sin^2(\sqrt{k} \chi) - \frac{kB_k^2}{A_k^2}}{\cos^2(\sqrt{k} \chi)}}\right),\label{eqA2a}\\
\text{for } k=0: & \int  \frac{d\chi}{\sqrt{A_k^2 - \frac{B_k^2}{\chi^2}}} = \frac{1}{|A_k|} \sqrt{\chi^2 - \frac{B_k^2}{A_k^2}},\label{eqA2b}\\
\text{for } k<0: & \int  \frac{d\chi}{\sqrt{A_k^2 + \frac{kB_k^2}{\sinh^2(\sqrt{-k}\chi)}}} = \frac{1}{|A_k| \sqrt{-k}} \text{arctanh}\left(\sqrt{\frac{\sinh^2(\sqrt{-k} \chi) + \frac{kB_k^2}{A_k^2}}{\cosh^2(\sqrt{-k} \chi)}}\right).\label{eqA2c}
\end{align}
\end{subequations}
These integrals can be combined using the $S_k$, $C_k$, and $T_k^{-1}$ functions given in Eqs.~\eqref{eq21},~\eqref{eq23}, and~\eqref{eq26}, respectively. We obtain
\begin{equation}
\int  \frac{d\chi}{\sqrt{A_k^2 - \frac{B_k^2}{S_k^2(\chi)}}} = \frac{1}{|A_k|} T_k^{-1}\left(\sqrt{\frac{S_k^2(\chi) - \frac{B_k^2}{A_k^2}}{C_k^2(\chi)}}\right).\label{eqA3}
\end{equation}
By using this integral along with the identities provided in Eqs.~\eqref{eq27a} and~\eqref{eq27b}, Eq.~\eqref{eqA1} yields
\begin{equation}
\chi(t) = S_k^{-1}\left(\sqrt{\left(1 - \frac{kB_k^2}{A_k^2}\right) S_k^2(R_k(t)) + \frac{B_k^2}{A_k^2}}\right).\label{eqA4}
\end{equation}
Here, the function $R_k$ has the following form
\begin{equation}
R_k(t) = \int_{t_i}^{t} \frac{\text{sgn}(\dot{\chi}(t')) |A_k| dt'}{a(t') \sqrt{a^2(t') + A_k^2}} + T_k^{-1}\left(\sqrt{\frac{S_k^2(\chi_i) - \frac{B_k^2}{A_k^2}}{C_k^2(\chi_i)}}\right).\label{eqA5}
\end{equation}
Addressing the integral terms in $R_k$ presents a non-trivial challenge, mainly due to the complexities involved in dealing with $\text{sgn}(\dot{\chi}(t'))$. By employing the method described in Appendix B of Ref.~\cite{Omar2}, we effectively address this issue and demonstrate that
\begin{equation}
R_k(t) = \int_{t_i}^{t} \frac{|A_k| dt'}{a(t') \sqrt{a^2(t') + A_k^2}} + \text{sgn}(\dot{\chi}_i) T_k^{-1}\left(\sqrt{\frac{S_k^2(\chi_i) - \frac{B_k^2}{A_k^2}}{C_k^2(\chi_i)}}\right).\label{eqA6}
\end{equation}
This proves the expression for the comoving radial solution $\chi(t)$ as outlined in Eq.~\eqref{eq45a}.
\section{Proving the Alternative Expression for the Angle}\label{A2}
Now, let us prove that the geodesic solution for the angle $\phi(t)$, as shown in Eq.~\eqref{eq45b}, is identical to the alternative geometrical formula presented in Eq.~\eqref{eq59}. Our aim is to show that $\phi(t)$ does not explicitly depend on the scale factor $a(t)$. By dividing Eqs.~\eqref{eq38b} and~\eqref{eq38a}, we can eliminate the explicit relation with the scale factor, resulting in the following expression:
\begin{equation}
\frac{\dot{\chi}(t)}{\dot{\phi}(t)} = \frac{\text{sgn}(\dot{\chi}(t)) }{B_k} S_k^2(\chi(t))\sqrt{A_k^2 - \frac{B_k^2}{S_k^2(\chi(t))}}.\label{eqB1}
\end{equation}
Upon separating variables $\phi$ and $t$, we arrive at
\begin{equation}
\int_{\phi_i}^{\phi} d\phi = B_k \int_{t_i}^{t} \frac{\text{sgn}(\dot{\chi}(t')) \dot{\chi}(t')}{S_k^2(\chi(t')) \sqrt{A_k^2 - \frac{B_k^2}{S_k^2(\chi(t'))}}} dt'.\label{eqB2}
\end{equation}
Applying the method from Appendix C of Ref.~\cite{Omar2} for the flat case to address the presence of $\text{sgn}(\dot{\chi}(t'))$ inside the integral, we find
\begin{equation}
\phi(t) = B_k \int_{\chi_i}^{\chi} \frac{\text{sgn}(\dot{\chi}) d\chi'}{S_k^2(\chi') \sqrt{A_k^2 - \frac{B_k^2}{S_k^2(\chi')}}} + \phi_i.\label{eqB3}
\end{equation}
The integral in the right-hand side of this equation can be performed for each specific geometrical space using the following formulae
\begin{subequations}
\begin{align}
\text{for } k>0: & \int \frac{kd\chi}{\sin^2(\sqrt{k} \chi) \sqrt{A_k^2 - \frac{kB_k^2}{\sin^2(\sqrt{k} \chi)}}} = \frac{1}{|B_k|} \arctan\left(\sqrt{\frac{A_k^2 \sin^2(\sqrt{k} \chi) - kB_k^2}{kB_k^2 \cos^2(\sqrt{k} \chi)}}\right),\label{eqB4a}\\
\text{for } k=0: & \int \frac{d\chi}{\chi^2 \sqrt{A_k^2 - \frac{B_k^2}{\chi^2}}} = \frac{1}{|B_k|} \arctan\left(\sqrt{\frac{A_k^2}{B_k^2} \chi^2 - 1}\right),\label{eqB4b}\\
\text{for } k<0: & \int \frac{-kd\chi}{\sinh^2(\sqrt{-k} \chi) \sqrt{A_k^2 + \frac{kB_k^2}{\sinh^2(\sqrt{-k} \chi)}}} = \frac{1}{|B_k|} \arctan\left(\sqrt{\frac{A_k^2 \sinh^2(\sqrt{-k} \chi) + kB_k^2}{-kB_k^2 \cosh^2(\sqrt{-k} \chi)}}\right).\label{eqB4c}
\end{align}
\end{subequations}
These integrals can be unified using the $S_k$ and $C_k$ from Eqs.~\eqref{eq21} and~\eqref{eq23}, respectively, as follows
{\small
\begin{align}
\int \frac{d\chi}{S_k^2(\chi) \sqrt{A_k^2 - \frac{B_k^2}{S_k^2(\chi)}}} = \frac{1}{|B_k|}\arctan\left(\sqrt{\frac{A_k^2 S_k^2(\chi) - B_k^2}{B_k^2 C_k^2(\chi)}}\right).\label{eqB5}
\end{align}}
By using this integral, the expression for $\phi(t)$ in Eq.~\eqref{eqB3} can be proved to have the following form
\begin{equation}
\phi(t) = \sin(B_k) \left[\text{sgn}(\dot{\chi}(t)) \arctan\left(\sqrt{\frac{A_k^2 S_k^2(\chi(t)) - B_k^2}{B_k^2 C_k^2(\chi(t))}}\right) - \text{sgn}(\dot{\chi}_i) \arctan\left(\sqrt{\frac{A_k^2 S_k^2(\chi_i) - B_k^2}{B_k^2 C_k^2(\chi_i)}}\right)\right] + \phi_i.\label{eqB6}
\end{equation}
This proves the alternative expression for the angle $\phi(t)$ as expressed in Eq.~\eqref{eq59}.
\section{Checking that the Solutions Satisfy the
Euler-Lagrange Equations}\label{A3}
We now confirm that our solution for the comoving radial-angular velocity in Eqs.~\eqref{eq38a} and~\eqref{eq38b} for generally curved spacetime satisfy the Euler-Lagrange equations~\eqref{eq33a} and~\eqref{eq33b}. Let’s begin with the first equation~\eqref{eq33a} by computing its left-hand and right-hand sides to get the following
{\small
\begin{subequations}
\begin{flalign}
&\text{LHS}=\frac{d}{dt} \left[ \frac{a^2 \dot{\chi}}{\sqrt{1 - a^2 \dot{\chi}^2 - a^2 S_k^2(\chi) \dot{\phi}^2}} \right] = \text{sgn}(\dot{\chi}) \frac{d}{dt} \left[ \sqrt{A_k^2 - \frac{B_k^2}{S_k^2(\chi)}} \right] =\text{sgn}(\dot{\chi}) \frac{\frac{B_k^2 \dot{\chi} C_k(\chi)}{S_k^3(\chi)}}{\sqrt{A_k^2 - \frac{B_k^2}{S_k^2(\chi}}} = \frac{1}{a} \frac{\frac{B_k^2 C_k(\chi)}{S_k^3(\chi)}}{ \sqrt{a^2 + A_k^2}}
,\label{eqC1a}&\\
&\text{RHS}=\frac{S_k(\chi) C_k(\chi) a^2 \dot{\phi}^2}{\sqrt{1 - a^2 \dot{\chi}^2 - a^2 S_k^2(\chi) \dot{\phi}^2}} = \frac{1}{a} \frac{\frac{B_k^2 C_k(\chi)}{S_k^3(\chi)}}{ \sqrt{a^2 + A_k^2}}.\label{eqC1b}&
\end{flalign}
\end{subequations}
}
This confirms that the comoving velocities in Eqs.~\eqref{eq38a} and~\eqref{eq38b} indeed satisfy the Euler-Lagrange equation~\eqref{eq33a}. Now, by doing the same thing for the Euler-Lagrange equation~\eqref{eq33b}, it can be checked easily from the constant of motion definition in Eq.~\eqref{eq35} that
\begin{equation}
\frac{d}{dt} \left[ \frac{a^2 S_k^2(\chi) \dot{\phi}}{\sqrt{1 - a^2 \dot{\chi}^2 - a^2 S_k^2(\chi) \dot{\phi}^2}} \right]=\frac{d}{dt}B_k=0,\label{eqC2}
\end{equation}
which confirms that our solution in Eqs.~\eqref{eq38a} and~\eqref{eq38b} also satisfies the Euler-Lagrange equation~\eqref{eq33b}. Consequently, both Euler-Lagrange equations~\eqref{eq33a} and~\eqref{eq33b} are checked to fulfill our solution indicated by Eqs.~\eqref{eq38a} and~\eqref{eq38b}.
\section{Summary: Explicit Geodesic Solutions for General Motion in Spherical and Hyperbolic Spaces}\label{A4}
The timelike geodesics for both radial and non-radial free motion in spatially spherical FLRW spacetime ($k>0$) are given by
{
\footnotesize
\begin{subequations}
\begin{flalign}
&\chi(t) = \frac{1}{\sqrt{k}} \arcsin\left(\sqrt{\left(1 - \frac{k B_k^2}{A_k^2}\right) \sin^2\left(\int_{t_i}^{t} \frac{\sqrt{k} |A_k| dt'}{a(t') \sqrt{a^2(t') + A_k^2}} + \text{sgn}(\dot{\chi}_i) \arctan\left(\sqrt{\frac{\sin^2(\sqrt{k} \chi_i) - \frac{k B_k^2}{A_k^2}}{\cos^2(\sqrt{k} \chi_i)}}\right)\right)} + \frac{k B_k^2}{A_k^2}\right),\label{eqD1a}\\
&\phi(t) = \sin(B_k) \left[\text{sgn}(\dot{\chi}(t)) \arctan\left(\sqrt{\frac{A_k^2 \sin^2(\sqrt{k} \chi(t)) - k B_k^2}{k B_k^2 \cos^2(\sqrt{k} \chi(t))}}\right) - \text{sgn}(\dot{\chi}_i) \arctan\left(\sqrt{\frac{A_k^2 \sin^2(\sqrt{k} \chi_i) - k B_k^2}{k B_k^2 \cos^2(\sqrt{k} \chi_i)}}\right)\right] + \phi_i.\label{eqD1b}
\end{flalign}
\end{subequations}}
The timelike geodesics for both radial and non-radial free motion in spatially hyperbolic FLRW spacetime ($k<0$) are given by
{
\footnotesize
\begin{subequations}
\begin{flalign}
&\chi(t) = \frac{1}{\sqrt{-k}} \text{arcsinh}\left(\sqrt{\left(1 + \frac{k B_k^2}{A_k^2}\right) \sinh^2\left(\int_{t_i}^{t} \frac{\sqrt{-k} |A_k| dt'}{a(t') \sqrt{a^2(t') + A_k^2}} + \text{sgn}(\dot{\chi}_i) \text{arctanh}\left(\sqrt{\frac{\sinh^2(\sqrt{-k} \chi_i) + \frac{k B_k^2}{A_k^2}}{\cosh^2(\sqrt{-k} \chi_i)}}\right)\right)} - \frac{k B_k^2}{A_k^2}\right),\label{eqD2a}\\
&\phi(t) = \sin(B_k) \left[\text{sgn}(\dot{\chi}(t)) \arctan\left(\sqrt{\frac{A_k^2 \sinh^2(\sqrt{-k} \chi(t)) + k B_k^2}{-k B_k^2 \cosh^2(\sqrt{-k} \chi(t))}}\right) - \text{sgn}(\dot{\chi}_i) \arctan\left(\sqrt{\frac{A_k^2 \sinh^2(\sqrt{-k} \chi_i) + k B_k^2}{-k B_k^2 \cosh^2(\sqrt{-k} \chi_i)}}\right)\right] + \phi_i.\label{eqD2b}
\end{flalign}
\end{subequations}
}
These explicit relations are given by the constants of motion $(A_k,B_k)$. However, we can expressed them in terms of $(v_{\chi,i},v_{\phi,i})$ by using Eqs.~\eqref{eq48}, and alternatively, in terms of $(v_{\text{pec},i},\psi_i)$ through Eqs.~\eqref{eq50}.
\nocite{*}

\bibliography{sn-bibliography}

\end{document}